\documentclass[aps]{revtex4}
\usepackage{epsfig,amsmath,amsfonts,wasysym}
\usepackage{graphicx}
\newcommand{\beao}{\begin{eqnarray*}}
\newcommand{\eeao}{\end{eqnarray*}}
\newcommand{\be}{\begin{equation}}\newcommand{\ee}{\end{equation}}
\newcommand{\bea}{\begin{eqnarray}}
\newcommand{\eea}{\end{eqnarray}}
\newcommand{\beq}{\begin{eqnarray}}
\newcommand{\eeq}{\end{eqnarray}}
\newcommand{\bs}{\begin{subequations}}
\newcommand{\es}{\end{subequations}}
\newcommand{\nn}{\nonumber}
\newcommand{\pa}{\partial}
\newcommand{\ep}{\epsilon}
\newcommand{\elm}{{electromagnetic }}

\newcommand{\om}{\omega}\newcommand{\Om}{\Omega}
\newcommand{\al}{\alpha}
\newcommand{\E}{{\mathbf E}}
\newcommand{\B}{{\mathbf B}}
\renewcommand{\r}{{\mathbf r}}

\newcommand{\Ref}[1]{(\ref{#1})}
\newcommand{\TE}{{\rm TE}}
\newcommand{\TM}{{\rm TM}}
\newcommand{\adb}{\allowdisplaybreaks }
\begin{document}
\title{On the vacuum energy of a spherical plasma shell}

\author{M. Bordag\footnote{bordag@itp.uni-leipzig.de} ${}^{a}$ and N.
Khusnutdinov\footnote{nrk@kazan-spu.ru}${}^{b}$}

\address{${}^a$Institute for Theoretical Physics, Leipzig University, Vor dem
Hospitaltore 1, D-04103 Leipzig \\${}^b$Department of Physics, Kazan State
University, Kremlevskaya 18, Kazan 420008, Russia, and \\ Department of Physics,
Tatar State University of Humanity and Education, \\ Tatarstan 2, Kazan 420021,
Russia}

\begin{abstract}
We consider the vacuum energy of the electromagnetic field interacting with a spherical plasma shell together with a model for the classical motion of the shell. We calculate the heat kernel coefficients, especially that for the TM mode, and carry out the renormalization by redefining the parameters of the classical model. It turns out that this is possible and results in a model, which in the limit of the plasma   shell becoming an ideal conductor reproduces the vacuum energy found by Boyer in 1968.
\end{abstract}
\maketitle
%\newpage
%
\section{Introduction}
The present paper is devoted to the discussion of the renormalization of vacuum
energy in the presence of boundaries or singular background fields in
application to the Casimir effect and it is aimed to partially fill the gap
between the two well understood situations. These are, on the one side, the
Casimir force between distinct objects which is always finite and, on the other
side, the vacuum energy in   smooth background fields which can be renormalized
by standard methods of quantum field theory. In between these two, the situation
is not finally settled. Especially in \cite{Graham:2003ib} it was questioned
whether boundaries can be incorporated at all into a well posed renormalization
program. For instance, it was argued that the process of making the background
field concentrated on a surface is not physical.

The aim of the present paper is to discuss an example of a background field
concentrated on a surface having both, a well posed renormalization procedure
for the vacuum energy and a meaningful physical interpretation. As model we take
a spherical plasma shell interacting with the \elm field and we allow for a
classical vibrational motion of the shell.
The investigation of the plasma shell model was pioneered by Barton \cite{BIII}
and it is aimed to describe the $\pi$-electrons in a C$_{60}$-molecule.

The heat kernel coefficients for such system are known to a large
extend. Since the polarizations for the \elm field separate into
the usual TE and TM modes, one is faced with two scalar problems,
where, however, the s-wave contribution must be dropped. For the
TE modes it is a delta function potential on the shell. The
corresponding heat kernel coefficients  (including the s-wave)
were first calculated in \cite{Bordag:1999vs}, later generalized
in \cite{Bordag:2004rx}, and the finite part of the vacuum energy
was calculated in \cite{Scandurra:1998xa}.   For the TM mode the
scalar problem corresponds to a $\delta'$-potential on the shell
and the corresponding heat kernel coefficients were calculated for
a plane shell only, \cite{Bordag:2005qv}. The problem with the TM
mode is that the corresponding spectral problem is not elliptic
and that the standard methods do not work. So, for example, for
the plane shell even the zeta function cannot be defined
\cite{Bordag:2005qv}. For the spherical shell, the zeta function
exists, but, as we will see below, it has double poles. It should
be mentioned that the $\delta'$-potential was considered in
\cite{Milton:2004ya} (where also the relevant literature was
collected), however with a coupling different from that following
within the plasma shell model (compare the Jost function in (4.41)
in \cite{Milton:2004ya} with \Ref{FTEM2} below).

In the present paper we calculate the heat kernel coefficients for the plasma
shell model, especially that for the TM modes. Using these, and the simplest
possible model for a classical motion of the shell, we construct a consistent
scheme for the renormalization. Within this scheme we define the renormalized
vacuum energy of the \elm field and calculate it numerically. Also we discuss
several limiting cases including the limit of the plasma shell becoming an ideal
conducting sphere.

Throughout the paper we use units with $\hbar=c=1$.

\section{The plasma shell model and its renormalization}\label{sec:2}
We consider the plasma shell model investigated, for example, in \cite{BIII} which is aimed to
model the $\pi$-electrons in a C$_{60}$-molecule. These electrons are described
by an electrically charged fluid whose motion is confined the shell. Further,
the model contains an immobile, overall electrically neutralizing background
aimed
to describe the carbon atoms and the remaining electrons.  The fluid is allowed
a non-relativistic motion. Of course, this model is a quite crude
simplification, especially because the motion of the electrons should rather
follow a relativistic dispersion relation \cite{Maksimenko2002,KN}. On the other
hand side it appears to be physically meaningful and should therefore result in
physically meaningful results for the vacuum energy. For instance, it should allow for a treatment of
the vacuum fluctuation of the \elm field coupled to the plasma shell.

The interaction of the plasma shell with the \elm field results in matching
conditions on the \elm field across the shell as shown in \cite{BIII} (and
earlier, for a plane sheet, in \cite{BI}). These conditions do not depend on the
state of the excitations of the fluid. The vacuum energy can be calculated from
the fluctuations of the \elm field whereas the fluctuations of the fluid must
not be taken into account as shown in \cite{BORDAG2007B} (or vice verse).
In this setup, the polarizations of the \elm field separate  into TE- and
TM-modes. For the electric and the
magnetic fields the corresponding mode expansions read
\bea
\E^{\rm TE}(t,\r)&=&
\sum_{{l\ge1\atop  |m|\le l}}\int_0^\infty \frac{d k}{\pi}
\frac{1}{\sqrt{2\om}}
\left(e^{-i\om t}
f_{l,m}(k,r) \mathbf{L} \frac{1}{\sqrt{L^2}}\, Y_{l,m}(\vartheta,\varphi)
+c.c. \right),
\nn \\
\B^{\rm TE}(t,\r)&=&   -\frac{i}{\sqrt{-\Delta}} \ \nabla \times \E^{\rm
TE}(t,\r),
\label{TE1}
\eea
where $\mathbf{L}$ is the orbital momentum operator and $ \omega=k$ follows from the wave equation . The radial
wave function
$f_{l,m}(kr)$ must be regular in the origin and across the shell it must fulfill
the matching conditions
\bea
\lim_{r\to R+0}f_{l,m}(kr)-\lim_{r\to R-0}f_{l,m}(kr)&=&0, \nn \\
\lim_{r\to R+0}(rf_{l,m}(k,r))'-\lim_{r\to R-0}(rf_{l,m}(k,r))'&=&\Om R
f_{l,m}(kR),\label{mcf}
\label{BC1}
\eea
where only the parameter
\be
\Om=\frac{4\pi ne^2}{mc^2}
\label{Om}
\ee
carries information on the properties of the fluid like its density $n$ and mass
$m$. It can be interpreted as a kind of plasma frequency in parallel to the
plasma frequency of a dielectric. For C$_{60}$ the corresponding wave lengths
is of the order of micrometers. The mode expansions for the TM polarization read
 by duality
\bea
\B^{\rm TM}(t,\r)&=&
\sum_{{l\ge1\atop  |m|\le l}}\int_0^\infty \frac{d k}{\pi}
\frac{1}{\sqrt{2\om}}
\left(e^{-i\om t}
 g_{l,m}(k,r) \mathbf{L} \frac{1}{\sqrt{L^2}}\, Y_{l,m}(\vartheta,\varphi)
+c.c. \right),
\nn \\
\E^{\rm TM}(t,\r)&=&   \frac{i}{\sqrt{-\Delta}} \ \nabla \times \B^{\rm
TM}(t,\r).
\label{TM1}
\eea
The matching conditions are different,
\bea
\lim_{r\to R+0}(rg_{l,m}(k,r))'-\lim_{r\to R-0}(rg_{l,m}(k,r))'&=&0, \nn \\
\lim_{r\to R+0}g_{l,m}(kr)-\lim_{r\to R-0}g_{l,m}(kr)&=&
-\frac{\Om}{k^2R}
(Rg_{l,m}(k,R))'.
\label{BC2}\label{mcg}
\eea
Considered as a scalar problem, the matching conditions \Ref{mcf} of the TE mode
are equivalent to a delta function potential $\Om \delta(r-R)$ in the wave
equation and the conditions \Ref{mcg} of the TM mode loosely speaking correspond
to the derivative of a delta function. A difference is that in the scalar
problems the zeroth orbital momentum, $l=0$, or s-wave contribution is present
whereas in the \elm case it is absent, i.e., the sums over $l$ in \Ref{TE1} and
in \Ref{TM1} start from $l=1$. In the limit $\Omega\to \infty$ which is formally the ideal conductor  limit
%\cite{Boy68}
the boundary conditions \Ref{mcf} and \Ref{mcg}
 became  Dirichlet boundary conditions for TE polarization and  Neumann
for TM polarization.

We extend this model by allowing for radial vibrations (breathing mode) of the
plasma shell. In C$_{60}$ these are determined by the elastic forces acting
between the carbon atoms. Without going here in any detail we describe these
vibrations phenomenologically by a Hamilton function
\be
H_{\rm class}=
\frac{p^2}{2m}+\frac{m}{2}\,\om_b^2\left(R-R_0\right)^2+E_{\rm rest}
\label{H1}
\ee
with a momentum $p=m\dot{R}$. Here $m$ is the mass of the shell, $\om_b$ is the
frequency of the breathing mode, $R_0$ is the radius at rest  and $E_{\rm rest}$
is the energy which is required to bring the pieces of the shell apart, i.e., it
is some kind of ionization energy.

Now we consider a system consisting of the classical motion of the shell as
described by $H_{\rm class}$ and the vacuum energy $E_{\rm vac}$ of the \elm
field interacting with the shell by means of the matching conditions \Ref{mcf}
and \Ref{mcg}. We assume the classical motion adiabatically slow such that the
vacuum energy can be taken as a function   of the mountainous radius of the
shell, $E_{\rm vac}=E_{\rm vac}(R)$, and we neglect the backreaction of the \elm
field on the shell. Under these assumptions the energy of the classical system,
$E_{\rm class}(R)=H_{\rm class}$, and the vacuum energy add up to the total
energy of the considered system,
\be
E_{\rm tot}=E_{\rm class}(R)+E_{\rm vac}(R).
\label{Etot}
\ee

Next we consider the ultraviolet divergences of the vacuum energy. These are
given in general terms by the heat kernel coefficients $a_n$ (we use the notations of
\cite{Bordag:2001qi} and we can define a 'divergent part' of the vacuum energy
which is, as known,  not uniquely defined. It depends on the kind of
regularization one has to introduce. For instance, in zeta functional
regularization, the regularized vacuum energy reads
\be
E_{\rm vac}(s)=\frac{\mu^{2s}}{2}\sum_n\om_n^{1-2s},
\label{Ezeta}
\ee
where $\mu$ is an arbitrary parameter with the dimension of a mass
and with a frequency damping function it is,
\be
E_{\rm vac}(\delta)=\frac{1}{2}\sum_n\om_n\,e^{-\delta\om_n},
\label{Edelta}
\ee
where $\om_n$ are the frequencies of the quantum fluctuations of the \elm field.
In our problem the spectrum is continuous, but for the moment it is more
instructive to keep the notations of a discrete spectrum. In zeta functional
regularization, the divergent part reads
\be
E_{\rm vac}^{\rm div}(s)=
-\frac{a_2}{32\pi^2} \left(\frac{1}{s}+\ln \mu^2\right),
\label{Edivs}
\ee
where we used the notations of \cite{Bordag:2001qi} in which the heat kernel expansion reads
\be
K(t)\sim\frac{1}{(4\pi t)^{3/2}}\left(a_0+a_{\frac12}\sqrt{t}+ a_1\,t+\dots\right)\,.
\label{hke}
\ee
In the scheme with the frequency damping we have
\be
E_{\rm vac}^{\rm div}(\delta)=
\frac{3a_0}{2\pi^{2}}{1\over \delta^{4}}
+\frac{a_{1/2}}{4{\pi^{3/2}}}{1\over \delta^{3}}
+\frac{{a}_{1}}{{8\pi^{2}}}{1\over \delta^{2}}
+ \frac{{a}_2}{16\pi^2} \ln \delta.
\label{Edivd}
\ee
The regularizations are removed by $s\to0$ resp. $\delta\to0$. These formulas
follow, for example, from section 3.4 in \cite{Bordag:2001qi} for $m=0$.

The idea of the renormalization is to have in the classical energy $E_{\rm
class}$ parameters which can be changed in a way to absorb $E_{\rm vac}^{\rm
div}$. In the considered model such parameters are the mass $m$ of the shell,
the frequency $\om_b$ of the breathing mode, the radius at rest $R_{0}$, and the
energy $E_{\rm rest}$. Now, whether this is possible, is a matter of the dependence of the heat kernel coefficients, especially
of $a_2$, on the radius $R$ which is the dynamical variable of the classical
system. In the considered, very simple model we have
only   a polynomial dependence on $R$ up to second order in \Ref{H1}. Since we
assumed adiabaticity for the motion of the shell we do not have a time
dependence in $a_2$ so that it cannot contain $\dot{R}$. Hence, the kinetic
energy remains unchanged and, together with it, the mass $m$. Only the remaining
parameters, $\om_b$, $R_0$ and $E_{\rm rest}$ can be used to accommodate the
divergent part. In fact, this turns out to be sufficient for the considered
model. As it will be seen below, the heat kernel coefficients $a_0,\dots,a_2$ which enter the
divergent part, depend on the radius polynomial and at most quadratically. In
this way this model is renormalizable.

It should be mentioned that this scheme is equivalent to the corresponding one
in quantum field theory with $E_{\rm vac}^{\rm div}$ in place of the
counterterms. Also the interpretation of the renormalization is similar. Namely,
we argue that the vacuum energy in fact cannot be switched off and what we
observe are   parameters like, for example in QED, electron mass and charge,
after renormalization.

Within this scheme of renormalization, the specific form of the heat kernel coefficients is
insignificant. The only what one has to bother of is its dependence on $R$ to fit
into the freedom of redefining the parameters in $E_{\rm class}$. If this is the
case, one may define a renormalized vacuum energy by means of
\be
E_{\rm vac}^{\rm ren}=
\lim_{s\to0}\left(E_{\rm vac}(s)-E_{\rm vac}^{\rm div}(s)\right)
\label{Erenvac}
\ee
(and the same with $\delta$ in place of $s$)
and one has now to consider
\be
E_{\rm tot}=E_{\rm class}+E_{\rm vac}^{\rm ren}
\label{Etotal}
\ee
in place of \Ref{Etot}. In this way, the question on how to remove the ultraviolet divergences is answered.

It remains, however, the question about the uniqueness of his procedure which comes in from
 the parameter $\mu$ in the zeta functional scheme or from the possibility of a
redefinition $\delta\to c\delta$ in the other scheme.

In the case of QED at this place one imposes   conditions on the analog of
$E_{\rm vac}^{\rm ren}$ in a way, the the mass and the charge take the values
one observes experimentally.

In our case a similar scheme is conceivable too. A different scheme, suggested
in \cite{Bordag:1999vs}, using the large mass expansion to fix the ambiguity
does not work here since the \elm field is massless. A way out could be to look
for a minimum of the total energy, $E_{\rm tot}$, \Ref{Etot}, which however
would imply to take the model \Ref{H1} seriously. This is not the aim of the
present paper. Instead,  as a normalization condition we demand that in the
limit  of the plasma frequency $\Om\to\infty$, where the matching conditions
\Ref{BC1} and \Ref{BC2} turn into that of an ideal conductor, we shall recover
the vacuum energy of a conducting spherical shell, i.e., just the quantity which
was first calculated by Boyer in \cite{BOYER1968}. Indeed, as we will see in the
next section, this is possible using the freedom of a finite renormalization.

\section{The Jost functions and the heat kernel coefficients for the spherical shell}

The \elm field interacting with the plasma shell is defined in the
whole space and it has a continuous spectrum. In that case the
vacuum energy, after the subtraction of the contribution of the
empty space, can be represented in the form (see Eq.(3.43) in
\cite{Bordag:2001qi})
\be
E_0(s)=-\frac{\cos \pi s}{\pi}
\mu^{2s}\sum_{l=1}^\infty\,\nu\int\limits_0^\infty dk\, k^{1-2s}\frac{\pa}{\pa
k} \ln f_l(ik)
\label{3E1}
\ee
with $\nu=l+1/2$. The   arbitrary parameter $\mu$ has the dimension of a mass and $f_l(k)$ is the Jost function of the corresponding scattering problem. It is determined by the wave equation and the matching conditions \Ref{BC1} and \Ref{BC2} respectively for the TE and TM modes. Here we have to consider the regular scattering solution which is defined as that solution which for $k\to 0$ turns into the free solution. For $r\to\infty$, it describes a superposition of incoming and outgoing spherical waves and  in our model it can be written in the form
\be
\phi^{\rm sc}_{l}(k,r)=j_l(kr)\Theta(R-r)
+\frac{1}{2}\left(f_l(k)h^{(2)}_l(kr)+f^*_l(k)h^{(1)}_l(kr)\right)\Theta(r-R),
\label{scsol}
\ee
where $j_l(x)=\sqrt{\pi/2x}J_{l+1/2}(x)$ and $h^{(1,2)}_l(x)=\sqrt{\pi/2x}H^{(1,2)}_{l+1/2}(x)$ are the spherical Bessel functions and $f_l(k)$ and $f^*_l(k)$ are the Jost function and its complex conjugate. For $r\ne R$ these are solutions of the radial wave equation. Imposing the matching conditions \Ref{BC1} and \Ref{BC2} on \Ref{scsol}, the Jost functions can be determined separately for each polarization,
\bea
f^{\rm TE}_l(k)&=&1-i\Om k R^2 j_l(kR)h^{(1)}_l(kR) ,
\nn \\
f^{\rm TM}_l(k)&=&1+i\frac{\Om}{k}j'_l(kR){h^{(1)}_l}'(kR) .
\label{FTEM1}
\eea
%
%These are related by means of  $\delta_l(k)=(1/2i)\ln f_l(k)/f^*_l(k)$ in the usual way with the scattering phase shifts which were considered in \cite{BIII}.
The corresponding formulas for imaginary argument read
\bea
f^{\rm TE}_l(ik)&=&1+ \frac{\Om}{k}   \, s_l(kR)e_l(kR),
\nn \\
f^{\rm TM}_l(ik)&=&k^2\left(1- \frac{\Om}{k}   \, s'_l(kR)e'_l(kR) \right),
\label{FTEM2}
\eea
where we used the modified Riccati-Bessel functions
\be
s_l(x)=\sqrt{\frac{\pi x}{2}}\,I_{l+1/2}(x),\qquad
e_l(x)=\sqrt{\frac{2x}{\pi}}\,K_{l+1/2}(x).
\ee
In \Ref{FTEM1} and \Ref{FTEM2} we made use of the freedom to multiply the Jost functions by a constant which does not influence the vacuum energy \Ref{3E1}.

In zetafunctional regularization, the ultraviolet divergences manifest
themselves as poles in the of the regularized energy $E_0(s)$, \Ref{3E1}. In our
case the pole structure reads
\be
2\mu^{-2s}(4\pi)^{3/2}\Gamma\left(s-\frac12\right)E_0(s)=
\sum_{k\ge 0}\frac{a_{k/2}}{s- 2 + \frac{k}{2}}
+ \sum_{k\ge 3}\frac{a'_{k/2}}{\left(s-2 + \frac{k}{2}\right)^2}
+\dots \, .
\label{pols}\ee
and included the double poles which will appear below in the TM mode.
Fortunately, the  double poles start from $k=5$ and do not influence the
renormalization.

In order to find the  poles one has to construct the analytic continuation of $E_0(s)$ into the region where the sum and the integral in representation \Ref{3E1} do not converge. For this one may use the uniform asymptotic expansion $f^{\rm as}_l(ik)$ of the Jost function for large both, $\nu$ and $k$ with $z\equiv\frac{k}{\nu}$ fixed. We define
\be
E^{\rm as}_0(s)=-\frac{\cos \pi s}{\pi}
\mu^{2s}\sum_{l=1}^\infty\,\nu\int\limits_0^\infty dk\, k^{1-2s}\frac{\pa}{\pa
k} \ln f^{\rm as}_l(ik),
\label{3E2}
\ee
whose pole contributions coincide with that of $E_0(s)$, \Ref{3E1}.
In the following subsections we obtain the heat kernel coefficients separately for the TE and TM
modes. As for the TE modes the procedure is well known. One simply inserts the
uniform asymptotic expansions of the Bessel functions entering \Ref{FTEM2} and
the analytic continuation is an easy task. For the TM mode, however, this does
not work and a more sophisticated treatment is in order.

\subsection{The asymptotic expansion for the TE mode}
Directly inserting the known uniform asymptotic expansions of the Bessel
modified functions \cite{abra70b} into \Ref{FTEM2} one obtains with $k=\nu z$
\be
 f^{\TE}(ik) \simeq 1 +
 \frac{\Om R t}{2 \nu }
 \left(1 + \sum_{j\ge 1}  \frac{c^{\TE}_j}{\nu^{2j}}\right).
\label{fTEas1}
\ee
The $c^{\TE}_j$ are polynomials in $t=1/\sqrt{1+z^2}$ (see \cite{BIII}, Appendix A) and we used
\be
c^{\TE}_1=\frac{t^2}{8}\left(1-6t^2+5t^4\right),
\quad c^{\TE}_2=\frac{t^4(1-t^2)}{128}\left(27-553t^2+1617t^4-1155t^6\right).
\ee
Using this expansion we define the asymptotic part of the logarithm of the Jost function,
\be
\ln f_l^{\TE, as}(ik)=\sum_{i=1}^3\frac{D_i}{\nu^{i}},
\label{lnfasTE}
\ee
with
\bea
D_1&=&  \frac{\Omega R t}{2}, \nn \\
D_2&=&  -\frac{(\Omega R)^2t^2}{8}, \\
D_3&=& \frac{(\Omega R)^3 t^3}{24} + \frac{\Omega R t^3}{16} \left(1-5
t^2\right)
\left(1-t^2\right)\nn .
\label{DTE}\eea
Inserting this into \Ref{3E2} we define $E^{\rm TE,\,   as}_0(s)$
\be
E^{\rm TE,\, as}_0(s)=-\frac{\cos \pi s}{\pi}\mu^{2s}
\sum_{l=1}^\infty\,\nu^{2-2s} \,\int\limits_0^\infty dz\, z^{1-2s}\frac{\pa}{\pa
z} \ln f_l^{\TE, as}(ik).
\label{3E3}
\ee
This expression is in a suitable form for the analytic continuation because it can be immediately expressed in terms of known functions.
In \Ref{3E3}, the sum over $\nu$ results in Hurwitz zeta functions,
\be
\sum_{l=0}^{\infty}\nu^{-s}=\zeta_H\left(s;\frac{1}{2}\right)
\label{sumsc}
\ee
for the scalar field with the s-wave included and, without the s-wave,
\be
\sum_{l=1}^{\infty}\nu^{-s}=\zeta_H\left(s;\frac{3}{2}\right)
\label{sumem}
\ee
for the \elm field. The integration over $z$ can be carried out using
\be
 \int_0^\infty z^{1-2s} t^n dz =
 \frac{\Gamma (1-s)\Gamma
(s-1+\frac{n}{2})}{2\Gamma (\frac{n}{2})}\
\qquad \left(1 - \frac{n}{2}<s<1\right).
\label{auxint}
\ee
From Eq. \Ref{pols} we get then the heat kernel coefficients for the TE mode. These are shown in
Table \ref{table:TE}. We remind that for the TE polarization there are no double
poles which can be confirmed by direct inspection in the above formulas. This holds also if higher orders of the asymptotic expansion  from \Ref{fTEas1} are included. For the
coefficients for the scalar field, i.e., including the s-wave, we confirm the
results found in \cite{Bordag:1999vs,Vassilevich:2003xt}. The coefficients for
the \elm field become different starting from $k=2$.

\begin{table}[ht]
\begin{tabular}{|c|c|c|}\hline\hline
$k$ & $l=0,1,\ldots$ & $l=1,2,\ldots$ \\
\hline $0$ & 0 & 0\\
\hline $1/2$ & 0 & 0\\
\hline $1$ & $ -4\pi \Omega R^2$ & $-4\pi \Omega R^2$\\
\hline $3/2$ & $\pi^{3/2}\Omega^2R^2$ & $\pi^{3/2}\Omega^2R^2$\\
\hline $2$ & $ -\frac{2}{3}\pi \Omega^3 R^2$ & $-\frac{2}{3}\pi \Omega^3 R^2 +
{4\pi \Omega}$\\ \hline\hline
\end{tabular}
\caption{The first few heat kernel coefficients $a^{\rm TE}_k$ for the TE polarization, including the s-wave in the second column and without it in the third column.}
\label{table:TE}\end{table}

\subsection{The asymptotic expansion for the TM mode}
As before in the preceding subsection we insert the asymptotic expansions of
the Bessel functions into the Jost function \Ref{FTEM2} and obtain
\be
 f^{\rm TM}_l(ik) = z^2 +  \frac{\Om R  }{2 \nu t}
 \left(1 + \sum_{j\ge 1}  \frac{c^{\TM}_j}{\nu^{2j}}\right) .
\label{fTMas1}
\ee
The coefficients $c^{\TM}_j$ are  also polynomials in $t$ and we need only
\be
c^{\TM}_1=-\frac{t^2}{8}\left(1-6t^2+7t^4\right).
\ee
The direct insertion of this asymptotic expansion into $\ln f^{\rm TM}_l(ik)$
would produce powers of $z$ in the denominator and a term-by-term integration over $z$ would be impossible. Therefore we define
\be
p=z^2+\frac{\Om R}{2\nu t}
\label{p}
\ee
and perform the asymptotic expansion of the logarithm for large $\nu$ with
fixed $p$,
\be
\ln f_l^{\rm TM}(ik)   \sim
\ln p+
\ln \left(
        1+\frac{\Om R}{2\nu t p}\sum_{i\ge 1}\frac{c^{\TM}_j}{\nu^{2j}}
    \right) ,
\label{lnexp}
\ee
and define a part of this expansion as
\be
\ln f_l^{\rm TM,\, as}(ik)   =
\ln p + \frac{\Om R}{2t p}\frac{c^{\TM}_1}{\nu^{3}}.
\label{lnfasTM}
\ee
In fact, this is a partial re-summation of the expansion one would obtain acting
in the same way as in the preceding subsection. Obviously, the pole
contributions to $E_0^{\rm TM}(s)$ can be obtained from this expansion too. The
advantage of \Ref{lnfasTM} is besides allowing for a term-by-term integration
that it turns into the asymptotic expansion for Neumann boundary conditions in
the formal limit $\Om\to\infty$.

Inserting now \Ref{lnfasTM} into \Ref{3E2} we obtain for the asymptotic part of
the vacuum energy of the TM mode the expression
\be
E_0^{\rm TM,\, as}(s)=\mu^{2s}\sum_{l\geq 1}
\sum_{k=0}^{7}\sum_{n=1}^2
\sum_{r=0}^{4} Y_{k,n}^r\nu^{2-2s-r}I_{k,n}^{s}\left(\frac{\Om
R}{2\nu}\right)
\label{3E4}
\ee
with
\be
I_{k,n}^s(\alpha)=-\frac{\cos\pi s}{\pi}\int\limits_0^\infty dz \,
\frac{z^{2-2s}(1+z^2)^{-k/2}}{\left(z^2+\alpha\sqrt{1+z^2}\right)^n}
\label{int2}
\ee
and the non-zero coefficients $Y_{k,n}^r$ are given in Eqs. \Ref{Y}:
\be
\begin{array}{rclcrclcrcl}
 Y^0_{0,1} &=& 1,& \quad Y^1_{1,1} &=& \frac{Q}{4} ,
 \\ [5pt]
 Y^3_{3,1} &=& \frac{Q}{32},& Y^3_{5,1} &=& -\frac{9 Q}{16},& \quad Y^3_{7,1} &=& \frac{35
Q}{32},
\\[5pt]
Y^3_{1,2} &=& \frac{Q}{16},& Y^3_{3,2} &=& -\frac{3 Q}{8},& Y^3_{5,2} &=& \frac{7
Q}{16},\label{Y}
\\[5pt]
Y^4_{2,2} &=& \frac{Q^2}{64},& Y^4_{4,1} &=& -\frac{3 Q^2}{32},& Y^4_{6,2} &=&
\frac{7Q^2}{64},
\end{array}
\ee
with $Q = \Omega R$. As a result of the re-summations these integrals are more
complicated than that which appeared for the TE modes.
%At this step it is easy
%to verify that in the limit $\Omega \to \infty$ the above expression is finite
%and the Neumann boundary condition is re-obtained. Indeed, in this limit we obtain
%\begin{displaymath}
%\left. E_0^{\rm TM,\, as}(s)\right|_{Q\to\infty} = -\frac{\cos\pi s}{\pi} \sum_{l=1}^\infty
%\nu^{2-2s} \int_0^\infty dz z^{2-2s}\left\{\frac{t^2}{2}+ \frac{t^4 (1
%- 12t^2 + 21 t^4)}{8 \nu^2}\right\}
%\end{displaymath}
%which coincides with the corresponding expression for Neumann boundary condition.

In order to perform in \Ref{3E4} the analytic continuation in $s$ we would like
to expand the integrals \Ref{int2} in a series in powers of $\alpha$. The direct
expansion in the integrand is impossible because of the behavior for small $z$.
Therefore the idea is to move the integration contour away from passing through
$z=0$.

Splitting the cosine into two exponentials we   change the integration in the second part by the substitution $z\to -z$. Both part can be united into one integral over the whole axis and \Ref{int2} becomes
\be
I^s_{k,n}(\alpha)=-\frac{e^{i\pi s}}{2\pi}\int\limits_{-\infty}^\infty dz \,
\frac{z^{2-2s}(1+z^2)^{-k/2}}{\left(z^2+\alpha\sqrt{1+z^2}\right)^n}\,.
\label{int3}
\ee
Next we move the integration path into the upper half plane. There is a pole of
$n$-th order in $i z_0$. We move the integration path across this pole  which
gives an additional contribution which results in  the second terms in
\Ref{int3a}, \Ref{int3b} and \Ref{I01} below. Moving the path further upwards
we hit the branch cut starting from  $z=i$ which originates from $\sqrt{1+z^2}$
in \Ref{int3}. To handle the singular behavior in $z=i$ we  divide the contour
into two
parts, one is a circle around $z=i$ with a small  radius $\epsilon $ and
the other is the path closed to the two branches  of the cut with $z=ix$, $ x=1+\epsilon\dots\infty$.
Integrating in this integral by parts $k-2$ times, the surface terms  just  cancel  the divergent terms coming from the circle around $z=i$ and in the limit $\ep\to0$ we
obtain an integral (first terms) and an explicit  contribution (last terms) in Eqs.
\Ref{int3a} and \Ref{int3b},
%
%\bs\label{Integrals}
\bea
I_{2k+1,n}^s  (\alpha)&=& \frac{(-2)^{k+1}}{4\pi
(2k-1)!!} \int_1^\infty dx \sqrt{x-1} F^{(k+1)}(x)\label{int3a}  \\
 &&+ \frac{i^{2 s-1}}{(n-1)!}\left[ \frac{(z-iz_0)^n z^{2-2s}t^{2k+1}}{(z^2 +
\alpha\sqrt{z^2+1})^n}\right]^{(n-1)}_{z=iz_{0}}  \nonumber \\
&&+ \frac{i^{2s-1}}{(2k-1)!} \left[ \frac{(i+z^2)^{2-2s} }{((i+z^2)^{2} + \alpha
z \sqrt{2i + z})^n} \frac{1}{(2i+z^2)^{k+1/2}}\right]^{(2k-1)}_{z=0},
\nonumber%\adb
\eea
and
\bea
I_{2k,n}^s (\alpha)&=& -\frac{(-2)^{k}}{4\pi
(2k-3)!!}\int_1^\infty dx \sqrt{x-1} \Psi^{(k)}(x)\label{int3b} \\
 &&+ \frac{i^{2 s-1}}{(n-1)!} \left[
\frac{(z-iz_0)^n z^{2-2s}t^{2k}}{(z^2 + \alpha
\sqrt{z^2+1})^n}\right]^{(n-1)}_{z=iz_{0}}  \nonumber \\
&&+ \frac{i^{2 s-1}}{(2k-2)!} \left[ \frac{(i+z^2)^{2-2s} }{((i+z^2)^{2} +
\alpha z \sqrt{2i + z})^n} \frac{1}{(2i+z^2)^{k}}\right]^{(2k-2)}_{z=0},
\nonumber
\end{eqnarray}
%\es
where we introduced the notations
\begin{eqnarray*}
 z_0 &=& \sqrt{-\frac{\alpha^2}{2} + \frac{\alpha}{2} \sqrt{\alpha^2 +1}
},\adb\\
F(x) &=& x^{-s} \left[ \frac{1}{(-x - i \alpha \sqrt{x-1})^n } +
\frac{1}{(-x +
i \alpha \sqrt{x-1})^n }\right],\adb\\
\Psi(x) &=& \frac{x^{-s} }{i\sqrt{x-1}}\left[ \frac{1}{(-x - i \alpha
\sqrt{x-1})^n } - \frac{1}{(-x +
i \alpha \sqrt{x-1})^n }\right].
\end{eqnarray*}
These formulas hold for $k\ge0$ whereby for $k=0$  the last term in
$I^s_{2k,n}(\al)$ must be dropped. As an example  we note for the simplest case
with $k=0$ and $n=1$,
\begin{equation}
 I_{0,1}^s (\alpha) = \frac{\alpha}{2\pi} \int_1^\infty
\frac{dx x^{-s + \frac12}\sqrt{x-1} }{x^2 + \alpha^2 (x-1)} +
\frac{2^{s-\frac32} \alpha^{-s + \frac12} }{\sqrt{\alpha^2 + 4}} \left( -\alpha
+ \sqrt{\alpha^2 + 4} \right)^{-s+\frac32}.
\label{I01}
\end{equation}

The merit of the representations \Ref{int3a} and \Ref{int3b} is that these can
be directly expanded into powers of $\alpha$. For instance, from \Ref{I01} we
get
\begin{eqnarray}
I_{0,1}^s &=& \alpha^{-s+\frac12}\left[\frac{1}{2} + \frac{1}{4} (s-\frac32)
\alpha + \frac{1}{16} (s-\frac52) (s-\frac12) \alpha^2 + \frac{1}{96}
(s-\frac72) (s-\frac32) (s+\frac12) \alpha^3 + \ldots
 \right]\adb\nonumber\\
&&+ \frac{\alpha}{4 \sqrt{\pi }}\left[\frac{\Gamma
   (s)}{\Gamma (s+\frac32)} - \frac{3 \alpha^2}{2}\frac{\Gamma
\left(s+1\right)}{\Gamma (s+\frac72)} + \ldots\right].
\end{eqnarray}
Now we insert these expansions into $E_0^{\rm TM,\, as}(s)$, \Ref{3E4} with
$\alpha=\Om R/(2\nu)$. There the sum over the orbital momentum $\nu=l+1/2$
delivers directly Hurwitz zeta functions, $\zeta_H(a,3/2)$, with corresponding
$a$. In case the s-wave is included, the result would be expressed in terms
$\zeta_H (a,1/2)$.
Keeping the necessary number of contributions we come to
\begin{gather}\label{3E4_1}
E^{\rm TM,\, as}_0(s)=  \frac{4(\mu R)^{2s}}R\left\{-\frac{Q (s-\frac12) \Gamma
\left(s\right)}{8
\sqrt{\pi }\Gamma (s+\frac32)}\zeta_H(2s-1,\frac{3}{2})\adb\right.\\
+\left. \frac{Q (s-\frac12) \left(3 Q^2 - (2s+3) (2s+5) (
(2s-1) (7 s+\frac{31}2)+27)\right)
   \Gamma \left(s+1\right)}{192 \sqrt{\pi } \Gamma
   (s+\frac72)}\zeta_H(2s+1,\frac{3}{2}) + \dots \nonumber\adb \right.\\
+\left. 2^{s-\frac32} Q^{-s+\frac12}\left[\zeta_H(s-\frac32,\frac{3}{2})+\frac{Q
(s-\frac12)}{4}\zeta_H(s-\frac12,\frac{3}{2}) + \frac{\left((s-\frac12) Q^2 +
8\right)(s-\frac12)}{32}
\zeta_H(s+\frac12,\frac{3}{2})\right.\adb\nonumber\right.\\
+\left.\left. \frac{Q \left(\left((s-\frac12)^2-1\right) Q^2 + 24
(s+\frac{17}2)\right)s}{384}\zeta_H(s+\frac32,\frac{3}{2}) + \ldots \right]
\right\}. \nonumber
\end{gather}
From this representation and together with Eq. \Ref{pols} we obtain the   heat kernel
coefficients for the TM polarization. These are shown in   Table \ref{table:TM}.

These coefficients can be compared with that obtained for a plane plasma sheet
by  dividing by the area of the sphere, $4\pi R^2$, and taking the limit
$R\to\infty$. In fact, these coincide for $k\ge1$ with that obtained in
\cite{Bordag:2005qv}.

\begin{table}[ht]
\begin{tabular}{|c|c|c|}\hline\hline
$a_k^{\TM}$ & $l=0,1,\ldots$ & $l=1,2,\ldots$ \\
\hline $0$ & 0 & 0\\
\hline $1/2$ & $ 8\pi^{3/2} R^2$ & $ 8\pi^{3/2} R^2$\\
\hline $1$ & $ -\frac{4\pi}{3} \Omega R^2$ & $ -\frac{4\pi}{3} \Omega R^2$\\
\hline $3/2$ & $\frac{14}{3}\pi^{3/2}$ & $-\frac{10}{3}\pi^{3/2}$\\
\hline $2$ & $ -8\pi \Omega +\frac{2\pi}{15}\Omega^3 R^2$ & $ -4\pi \Omega
+\frac{2\pi}{15}\Omega^3 R^2$\\ \hline\hline
\end{tabular}\caption{The heat kernel coefficients for TM polarization. We
represent the calculation with $s$-wave (second column) and without $s$-wave
(third column). The difference appears starting from $k=3/2$.}\label{table:TM}
\end{table}

From Eqs. \Ref{3E4_1} together with \Ref{pols} also the double poles mentioned
in the introduction follows. It comes  at $s=-1/2$ from the zeta function
$\zeta_H(s+\frac32;\frac32)$ in the last line in \Ref{3E4} and the gamma
function in the left hand side of Eq. \Ref{pols} and corresponds to heat kernel
coefficient $a'_{5/2}$. There is no double pole at point $s=1/2$ which
corresponds to $a'_{3/2}$ due to factor $(s-1/2)$ at
$\zeta_H(s+\frac12;\frac32)$ in \Ref{3E4_1}. Therefore the logarithmic
contributions for heat kernel expansion starts from $a'_{5/2}$.

\section{The renormalized vacuum energy}
The renormalization of the vacuum energy is given by Eqs. \Ref{Erenvac} and
\Ref{Edivs} with the heat kernel coefficients calculated in the preceding section. As discussed in
Sec. \ref{sec:2}, it remains to dispose of the freedom of a finite
renormalization. It turns out that it is possible to join this with the behavior
for large $\Om$, i.e. with the limit of the plasma sphere to become an ideal
conductor.

We start from dividing the regularized vacuum energy into two parts,
\be
E_{\rm vac}(s)=E_{\rm vac}^{\rm num}+E_{\rm vac}^{\rm as}(s),
\label{4E1}
\ee
where $E_{\rm vac}^{\rm as}$ is defined by Eq.\Ref{3E2}   and the 'numerical' parts of the energy are defined by
\be
E_{\rm vac}^{\rm num}=
-\frac{1}{\pi}
 \sum_{l=1}^\infty\,\nu\int\limits_0^\infty dk\, k\frac{\pa}{\pa
k}
\left( \ln f_l(ik) - \ln f_l^{\rm as}(ik)\right).
\label{4Enum}
\ee
In this way the asymptotic part of the logarithm of the Jost function, $\ln
f_l^{\rm as}(ik)$, is subtracted in $E_{\rm vac}^{\rm num}$ and added back in
$E_{\rm vac}^{\rm as}$. In general, the definition of $\ln f_l^{\rm as}(ik)$ is
not unique. The only one has to ensure in any case is that the sum and the
integral in \Ref{4Enum} do converge  if one puts $s=0$ there. We did this in
Eq.\Ref{4Enum}. This part is called 'numerical' since it allows for a direct
numerical evaluation.

In the preceding section we used the expressions \Ref{lnfasTE} and
\Ref{lnfasTM} for $\ln f_l^{\rm as}(ik)$ for the calculation of
the heat kernel coefficients. In this section we take the same
expression for the TM mode \Ref{lnfasTM} and different expression
for TE mode, Eq.\Ref{4lnfas} below. In $\ln f_l^{\rm TE, \,
as}(ik)$ we make a re-summation like that in $\ln f_l^{\rm
TM,\,as}(ik)$, \Ref{lnfasTM}, in order to archive also for $\ln
f_l^{\rm TE, \, as}(ik)$ the property in the formal limit
$\Om\to\infty$ to turn into the corresponding expression for an
ideal conductor.

With this motivation we define
\be
\ln f_l^{\rm TE,\, as}(ik)   =
\ln w + \frac{\Om R t}{2 w}\frac{c^{\TE}_1}{\nu^{3}},
\label{4lnfas}
\ee
where
\be w=1+ \frac{\Om R t}{2 \nu}.\ee
We would like to stress again that a redefinition of $\ln f_l^{\rm as}(ik)$ is
 merely a redistribution of contributions between $E_{\rm vac}^{\rm num}$ and $
E_{\rm vac}^{\rm as}(s)$. With the definitions \Ref{4lnfas} and
\Ref{lnfasTM} we archived that $E_{\rm vac}^{\rm num}$ in the
limit $\Om\to\infty$ must turn into the corresponding ideal
conductor expressions. This follows because in \Ref{4Enum} both,
the Jost function and the part \Ref{4lnfas} of its asymptotic
expansion, do that and because in addition the integral and the
sum are convergent. Indeed, from the numerical evaluation we got
\bea\label{TENumLim}
 \lim_{Q\to \infty}E_{\rm vac}^{\rm TE,\,num} &=& \frac{0.00090282}{R},
\nn \\
 \lim_{Q\to \infty}E_{\rm vac}^{\rm TM,\,num}
 &=&\frac{-0.00160178}{R},
\eea
which is the same as if one takes conductor boundary conditions
from the beginning.  With the other part, $E_{\rm vac}^{\rm
as}(s)$, this is not such simple. The corresponding calculations
are carried out in the Appendix and the result is
\bea E_{\rm vac}^{\rm TE,\,as}(s) &=& -\frac{a_2^{\TE}
}{32\pi^{2}} \left[ \frac{1}{s} - 2 \ln\frac{\Omega}{2\mu}\right]
+ \frac{\Om^3R^2}{72\pi} + \frac{\Om}{180\pi} + E_{\rm vac}^{\rm
TE,\,an}+O(s),
\nn \\
E_{\rm vac}^{\rm TM,\,as}(s) &=& -\frac{a_2^{\TM}}{32\pi^{2}}
\left[ \frac{1}{s} - 2 \ln\frac{\Omega}{2\mu}\right] +
\frac{7\Om^3R^2}{1800\pi} - \frac{29\Om}{36\pi} + E_{\rm vac}^{\rm
TM,\,an}+O(s. \label{4Ean1} \eea
These expressions are sums of a divergent part (it is proportional to the heat
kernel coefficient $a_2$ as expected), two terms growing with $\Om$ and an
'analytical' part,
\bea
E_{\rm vac}^{\rm TE,\,an} &=& \sum_{l=1}^4 \mathcal{V}_l +
\sum_{l=1}^3 \widetilde{\mathcal{V}}_l,
\nn \\
E_{\rm vac}^{\rm TM,\,an} &=& \frac{1}{2}\mathcal{J}_1 +
\sum_{l=2}^6 \mathcal{J}_l + \sum_{l=1}^4 \widetilde{\mathcal{J}}_l.
\eea
All quantities entering $E_{\rm vac}^{\rm TE,\,an}$ and $E_{\rm vac}^{\rm TM,\,an}$ are defined in the Appendix. The analytical parts have   the 'necessary' limit for $\Om\to\infty$,
\bea \lim_{Q\to\infty} E_{\rm vac}^{\rm TE,\,an}&=&
\frac{17}{128R},
\nn \\
\lim_{Q\to\infty} E_{\rm vac}^{\rm TM,\,an}&=&-\frac{11}{128R}.
\label{4inf}\eea

With Eqs. \Ref{4Ean1} and the property \Ref{4inf} we have all
information we need to complete the renormalization. We remind the
discussion in Sec. \ref{sec:2} that all terms which are
proportional to $R$, $R^2$ or which do not depend on $R$ can be
removed by a redefinition of the parameters in the classical part.
In \Ref{4Ean1} this concerns all except  the last ones, $E_{\rm
vac}^{\rm TE,\,an}$ and $E_{\rm vac}^{\rm TM,\,an}$. That means,
that we not only can remove the contribution proportional to
$a_2$, but also the terms growing with $\Om$. For this reasons we
define the renormalized vacuum energies by
\bea
E_{\rm vac}^{\rm TE,\,ren}&=&E_{\rm vac}^{\rm TE,\,num}+E_{\rm vac}^{\rm TE,\,an},
\nn \\ [4pt]
E_{\rm vac}^{\rm TM,\,ren}&=&E_{\rm vac}^{\rm TM,\,num}+E_{\rm vac}^{\rm TM,\,an}.
\label{4Eren}\eea
With these formulas we completed the model consisting of the classical energy
and the vacuum energy which is the sum of the two contributions in \Ref{4Eren}.
The main merit of this vacuum energy is that it turns for $\Om\to\infty$ into
the ideal conductor limit. Using the formulas for the Jost functions and their
asymptotic parts and also the formulas in the Appendix, it is possible to
evaluate this vacuum energy numerically. The results are shown in the figures
\ref{ecasimir} and \ref{eqfix} for the dimensionless function $\mathcal{E}$
defined by
\be
E_{\rm vac}^{\rm ren} = \frac{\mathcal{E}(\Om R)}{R}
\ee
as functions of their arguments $x=\Om R$.

\begin{figure}[h]
\begin{center}
{\epsfxsize=7truecm\epsfbox{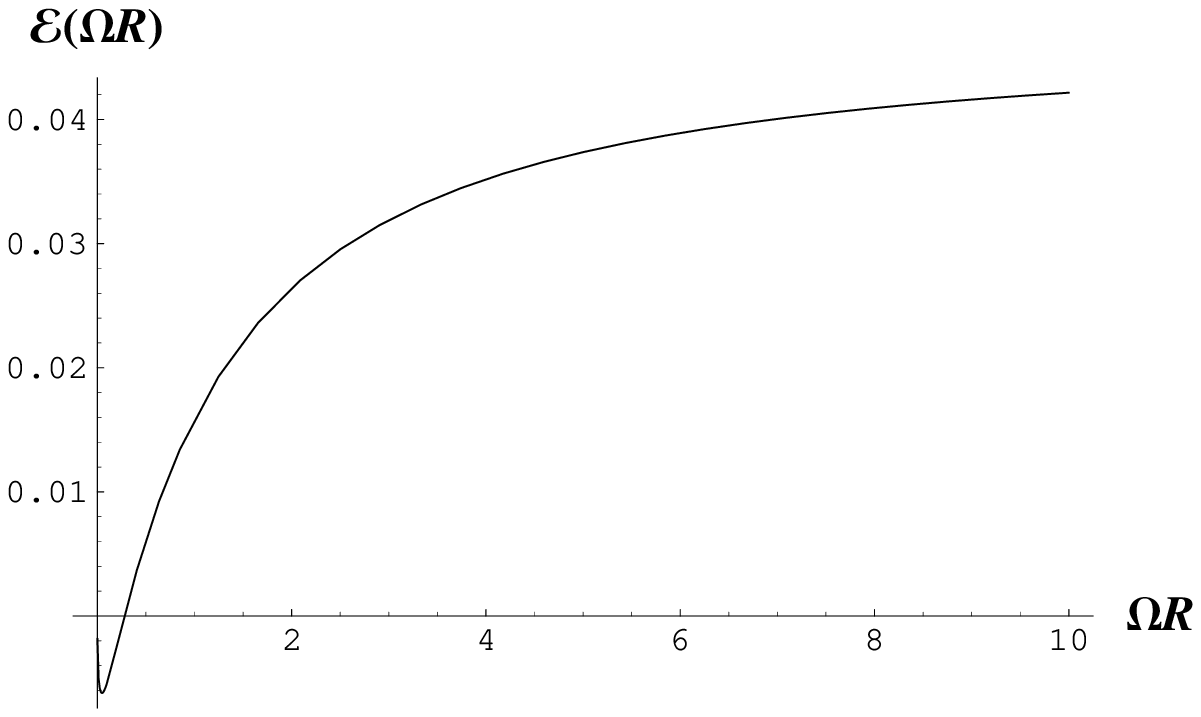}}%
{\epsfxsize=7truecm\epsfbox{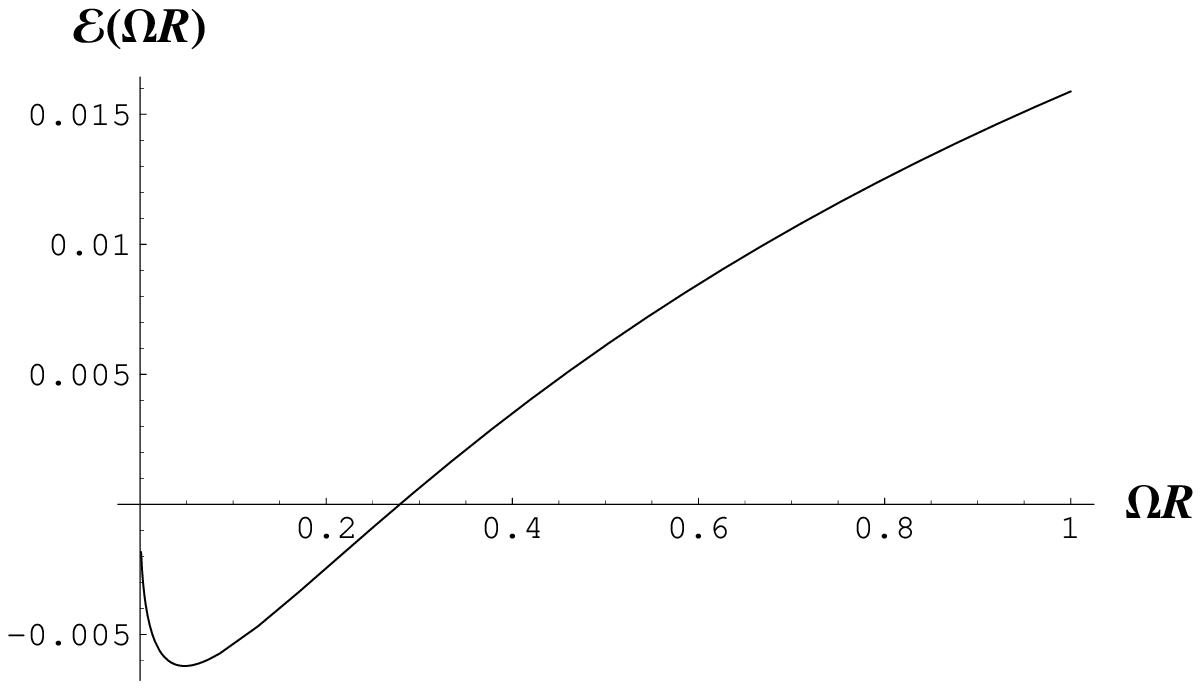}}
\end{center}
\caption{
The function $\mathcal{E}(\Om R)=  RE_{\rm vac}^{\rm ren}$ plotted as function
of $\Om$. For large $\Om$  it tends to the ideal conductor limit, $\lim_{\Om
R\to\infty}=0.0046 $ (left panel). For small $\Om$ (right panel)  it takes
negative values and decreases as $\mathcal{E}(\Om R)\sim-0.0589\sqrt{\Om R}$.
}
\end{figure}\label{ecasimir}

\begin{figure}[htb]
\begin{center}
{\epsfxsize=9truecm\epsfbox{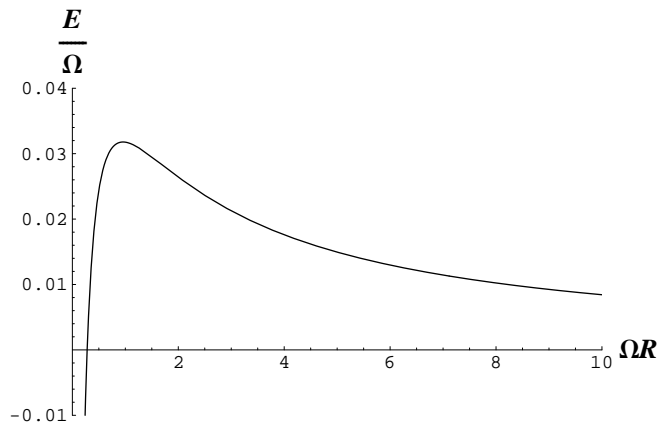}}
\end{center}
\caption{ The renormalized vacuum energy divided by $\Om$, $E_{\rm
vac}^{\rm ren}/ \Om$, as a function of $R$. For large $R$ it
approaches the ideal conductor limit and for small radii,  $R
\lesssim \Om^{-1}$, it becomes attractive.}
\end{figure}\label{eqfix}

It is also interesting to consider the limit of small argument of these
functions which is equivalent to a small plasma frequency $\Om$.
For the TE mode the main contributions come  from the integrals
\begin{eqnarray*}
{\mathcal{V}}_1(\Om R) &=&  \frac{\Om R\ln \Om R}{48\pi} + O(\Om R),\adb\\
{\mathcal{V}}_3(\Om R) &=&  \frac{\Om R\ln \Om R}{48\pi} + O(\Om R),\adb\\
\widetilde{\mathcal{V}}_1(\Om R) &=& -\frac{\Om R\ln \Om R}{4\pi} + O(\Om R).
\end{eqnarray*}
For the TM mode we get accordingly%
\begin{eqnarray*}
 \mathcal{J}_2(\Om R) &=& \frac{3(4-\sqrt{2})\zeta_R(\frac{5}{2})}{64\pi^2 \sqrt{2}}
\sqrt{\Om R} +  O(\Om R\ln \Om R),\adb\\
 \mathcal{J}_5(\Om R) &=& \sqrt{\Om R} \frac{(1-\sqrt{2})\zeta_R(\frac{1}{2})}{8 \sqrt{2}}
 + O(\Om R\ln \Om R),\adb\\
\widetilde{\mathcal{J}}_1(\Om R) &=& -\frac{\sqrt{\Om R}}{4} + O(\Om R\ln \Om R),\adb\\
\widetilde{\mathcal{J}}_3(\Om R) &=& \frac{\sqrt{\Om R}}{8} + O(\Om R\ln \Om R),
\end{eqnarray*}
where $\zeta_R(x)$ is the Riemann zeta function. The numerical parts are  $\sim
O(\Om)$ which is easy to show   using the next term of asymptotic expansion of the
Jost functions. In this way we obtain the following behavior of the vacuum energy for
small $  \Omega  $,
\begin{eqnarray}
 E^{\rm TE,\,ren}|_{\Om\to 0} &\simeq&   - \frac{5}{24\pi } \Omega \
\ln \Omega R,
\nn \\
 E^{\rm TM,\,ren}|_{\Om\to 0} &\simeq&
\left(\frac{3(4-\sqrt{2})\zeta(\frac{5}{2})}{64\pi^2 \sqrt{2}} +
\frac{(1-\sqrt{2})\zeta(\frac{1}{2})}{8\sqrt{2}} - \frac{1}{8}\right) \sqrt{\frac{\Omega}{R}}
\nn \\&&=
-0.0598 \sqrt{\frac{\Omega}{R}}.
\end{eqnarray}
Therefore the main contribution for the energy comes from the TM polarization.
The energy tends to zero proportional to $\sqrt{\Omega}$.  The same   behavior was observed in   Refs. \cite{Bordag:2005qv} and \cite{BORDAG2006F}.

\section{Conclusion}
In the foregoing sections we considered the vacuum energy of the \elm field interacting with a spherical plasma shell. We calculated the heat kernel coefficients for both polarizations. For the TE case the standard methods apply, for the TM case a re-expansion of the asymptotic expansion of the logarithm of the Jost function was helpful. It turned out that the vacuum energy in zeta functional regularization and, with it the corresponding zeta function, have double poles. This implies that the corresponding spectral problem is not elliptic. On the other hand, at least based on the calculations carried out in this paper, there is nothing which would diminish the reasonability of this model.

A basic concern of this paper is to construct a model allowing for
a physically meaningful interpretation of the renormalization. We
considered with the breathing mode of the shell the simplest model
for the classical motion of the shell. It turned out that this
model is  able to accommodate all renormalizations which we were
like to carry out. These are the removal of the pole in $s$, i.e.,
of the ultraviolet divergence, and the removal of all
contributions growing together with the plasma frequency $\Om$. It
should be mentioned that this includes also the removal of the
arbitrary constant $\mu$ which came in with the regularization.
The nontrivial statement which allowed for doing so is that the
dependence on the radius $R$ of all these contributions is
polynomial not exceeding $R^2$.

It would be  interesting to investigate the question whether this
procedure can be carried out also for more general deformations of
the shell. In principle, most ingredients for such a calculation
are available. Especially, the heat kernel coefficients for the TE
modes can be taken from \cite{Vassilevich:2003xt}. It would remain
to calculate the coefficients for the TM modes.

Concerning the arbitrariness of the normalization procedure we
would like to mention that the removal of the contributions
growing together with $\Om$  can be considered as a normalization
condition. It ensures the uniqueness of the renormalized vacuum
energy and makes this model physically meaningful. In this way,
the gap between the renormalization procedure in quantum field
theory in smooth background fields  and the removal of divergences
of the Casimir energy in the background of boundaries, as
suggested  for example in \cite{Blau:1988kv} (section 6.5), is
narrowed. At once in this way the much discussed vacuum  energy of
a conducting spherical shell now appears as a limiting case of a
slightly more physical model.

\section*{Acknowledgments}
The work of M.B. was supported by the research funding from the {EC's} Sixth
Framework Programme within the STRP project "PARNASS" (NMP4-CT-2005-01707).
N.K. is grateful to Institute of the Theoretical Physics of Leipzig
University where this work was done, for their hospitality. The work of NK was
supported by the DAAD program ``Mikhail Lomo\-nosov''. We acknowledge
valuable discussions with I. Pirozhenko and V. Nesterenko on the heat kernel
coefficients for the TM mode.
\section{Appendix}
In this appendix we perform the analytic continuation in the asymptotic parts of the vacuum energy defined in section IV. For that, we use the following  integral representations of some sums,
\begin{eqnarray}\label{AbelPlana}
\sum_{l = 0}\frac{\nu^{3-2s}}{\nu+a} &=& -\frac{\pi a^{3-2s}}{\sin
2\pi s} + 2\int_0^\infty \frac{dy y^{3-2s}}{y^2+a^2} \frac{-a\cos
\pi s + y \sin \pi s}{1+e^{2\pi y}}\adb,\ \frac{3}{2}< \Re s < 2,\\
\sum_{l = 0} \frac{\nu^{2-2s}}{\nu+a} &=& \frac{\pi a^{2-2s}}{\sin 2\pi
s} + 2\int_0^\infty \frac{dy y^{2-2s}}{y^2+a^2} \frac{a\sin \pi s+ y
\cos\pi s}{1+e^{2\pi y}},\ 1< \Re s < \frac{3}{2},\nonumber\\
\sum_{l = 0} \frac{\nu^{1-2s}}{\nu+a} &=& -\frac{\pi a^{1-2s}}{\sin 2\pi
s} - 2\int_0^\infty \frac{dy y^{1-2s}}{y^2+a^2} \frac{-a\cos \pi s+
y \sin\pi s}{1+e^{2\pi y}}\adb,\ \frac{1}{2}< \Re s < 1,\nonumber\\
\sum_{l = 0} \frac{\nu^{1-2s}}{(\nu+a)^2} &=& -\frac{2\pi (s-\frac12)
a^{-2s}}{\sin 2\pi s} - 2\int_0^\infty \frac{dy
y^{-2s}}{(y^2+a^2)^2} \frac{2ay\sin \pi s - (a^2-y^2)
\cos\pi s}{1+e^{2\pi y}}\adb, \ \Re s < 1,\nonumber\\
\sum_{l = 0} \frac{1}{(\nu+a)^2} &=& \pi a\int_0^\infty
\frac{dy}{y^2+a^2} \frac{1}{\cosh^2{\pi y}},\nonumber
\end{eqnarray}
which were obtained using the Abel-Plana formula in the form
\begin{displaymath}
 \sum_{l = 0}^\infty f(l+\frac{1}{2}) =
 \int_0^\infty dx \ f(x)
 -i \int_0^\infty dx \ \frac{f(iy)-f(-iy)}{1+e^{2\pi y}}.
\end{displaymath}

\subsection{TE polarization. $b = Qt/2$}
The expression for asymptotic part $E_{\rm vac}^{\rm TE,\,as}$ appearing from inserting \Ref{4lnfas} into \Ref{3E3}, after carrying out the differentiation, has the following form
\begin{eqnarray*}
E_{\rm vac}^{\rm TE,\,as} &=& \frac{\cos\pi s}{\pi R} (\mu R)^{2s}
\sum_{l=1}^\infty \nu^{2-2s} \int_0^\infty dz z^{2-2s}\label{Esing1}\\
&\times& \left\{ \frac{Qt^3}{2w\nu} +
  \frac{Qt^5\left( 3 - 30t^2 + 35t^4 \right)}{16w\nu^3} -
  \frac{Q^2t^6\left( 1 - 6t^2 + 5t^4 \right)}{32w^2  \nu^4}
\right\},\nonumber
\end{eqnarray*}
with $w = 1 + Qt/2\nu$ and $Q=\Om R$. We perform the calculations separately for the  contribution from each
power of $\nu$ using the formulas \Ref{AbelPlana} given above.

\begin{eqnarray*}
[\nu^{-1}]&:& \frac{\cos\pi s}{\pi } (\mu R)^{2s} \sum_{l=1}\nu^{2-2s}
\int_0^\infty dz z^{2-2s}\frac{Q t^3 }{2w\nu}\adb\\
&=&Q\frac{\cos\pi s}{2\pi } (\mu R)^{2s} \int_0^\infty dz
z^{2-2s}t^3\sum_{l=1}\frac{\nu^{2-2s}}{\nu + b}\adb\\
&=&\frac{1}{48\pi } \left( Q^3 - \frac{11 }{2}Q\right) \left[
\frac{1}{s} - 2\ln\frac{\Omega}{2\mu}\right] +
\frac{Q^3}{72\pi} + \mathcal{V}_1(Q) + \mathcal{V}_2(Q) +
\widetilde{\mathcal{V}}_1(Q),\\{}
[\nu^{-3}]&:& \frac{\cos\pi s}{\pi } (\mu R)^{2s} \sum_{l=1}\nu^{2-2s}
\int_0^\infty dz z^{2-2s}\frac{Q t^5\left( 3 - 30t^2 + 35t^4
\right)}{16w \nu^3}\adb\\
&=&Q\frac{\cos\pi s}{16\pi } (\mu R)^{2s} \int_0^\infty dz
z^{2-2s}t^5\left( 3 - 30t^2 + 35t^4
\right)\sum_{l=1}\frac{\nu^{-2s}}{\nu + b}\adb\\
&=&-\frac{Q}{96\pi } \left[ \frac{1}{s} -
2\ln\frac{\Omega}{2\mu}\right] + \frac{Q}{5040\pi} + \mathcal{V}_3(Q) +
\widetilde{\mathcal{V}}_2(Q),\\{}
[\nu^{-4}]&:& -\frac{\cos\pi s}{\pi } (\mu R)^{2s} \sum_{l=1}\nu^{2-2s}
\int_0^\infty dz z^{2-2s} \frac{Q^2 t^6\left( 1 - 6t^2 + 5t^4
\right){\epsilon }^4}{32 w^2}\adb\\
&=&-\frac{Q^2}{32\pi } \int_0^\infty dz z^{2}t^6\left( 1 - 6t^2 + 5t^4
\right)\sum_{l=1}\frac{1}{(\nu + b)^2}\adb\\
&=& \frac{3Q}{560\pi} + \mathcal{V}_4(Q) +
\widetilde{\mathcal{V}}_3(Q),
\end{eqnarray*}
where  the following integrals were introduced,
\begin{eqnarray*}
\mathcal{V}_1 &=& -\frac{Q}{2\pi}\int_0^\infty \frac{y dy}{1+e^{2\pi
y}} \ln\left[1+ \frac{4y^2}{Q^2}\right],\adb\\
\mathcal{V}_2 &=& -\frac{Q}{\pi}\int_0^\infty \frac{y^3
dy}{1+e^{2\pi y}} \int_0^\infty \frac{x  t^4 dx}{y^2 + Q^2 t^2/4}
\frac{1}{1+xt},\adb\\
\mathcal{V}_3 &=& \frac{Q}{8\pi}\int_0^\infty \frac{y dy}{1+e^{2\pi
y}} \int_0^\infty \frac{x^2  t^5 (3 - 30t^2 + 35t^4) dx}{y^2 + Q^2
t^2/4},\adb\\
\mathcal{V}_4 &=& \frac{Q}{16}\int_0^\infty \frac{y^2 dy}{\cosh^2\pi
y} \int_0^\infty \frac{x^2  t^5 (1 - 6t^2 + 5t^4) dx}{y^2 + Q^2
t^2/4}\adb\\
\widetilde{\mathcal{V}}_1(Q) &=& \frac{Q^2}{4\pi}\int_0^\infty
\frac{x^2t^4dx}{1+Qt}+ \frac{Q}{4\pi}(1- \ln 2Q)\adb\\
&=& \frac{1}{8\pi }\left(\pi - 4Q + i \pi {\sqrt{-1 + 4Q^2}} +
2{\sqrt{-1 + 4Q^2}}\textrm{arctanh}\left[\frac{2Q}{{\sqrt{-1 +
4Q^2}}}\right]\right)\adb\\
&+& \frac{Q}{4\pi}(1- \ln 2Q),\adb\\
\widetilde{\mathcal{V}}_2(Q)&=& -\frac{Q}{8\pi} \int_0^\infty dx x^2
t^5 \frac{3 - 30t^2 + 35t^4}{1+Qt},\adb\\
\widetilde{\mathcal{V}}_3(Q) &=& \frac{Q^2}{8\pi} \int_0^\infty dx
x^2 t^6 \frac{1 - 6t^2 + 5t^4}{(1+Qt)^2}.
\end{eqnarray*}
We note the following expressions which are necessary to consider the ideal conductor limit,
\begin{eqnarray*}
\lim_{Q\to \infty}\widetilde{\mathcal{V}}_1(Q) &=& \frac{1}{8},\\
\lim_{Q\to \infty}\widetilde{\mathcal{V}}_2(Q) &=& \frac{1}{256},\\
\lim_{Q\to \infty}\widetilde{\mathcal{V}}_3(Q) &=& \frac{1}{256}.
\end{eqnarray*}
All other integrals vanish in this limit.

\subsection{TM polarization. $a=Q/2tz^2$}
The expression for asymptotic part $E_{\rm vac}^{\rm TM,\,as}$ appearing from inserting \Ref{lnfasTM} into \Ref{3E3}, after carrying out the differentiation, has the following form
\begin{eqnarray*}
E_{\rm vac}^{\rm TM,\,as} &=& -\frac{2\cos\pi s}{\pi R} (\mu R)^{2s}
\sum_{l=1}^\infty \nu^{2-2s} \int_0^\infty dz
z^{2-2s}\adb \label{Esing}\\
&\times&\left\{\frac{1}{p}+ \frac{Q t}{4 p \nu } + \frac{Qt}{32 \nu ^3}
\left(\frac{\left(35 t^4-18 t^2+1\right) t^2}{p} + \frac{14 t^4-12
   t^2+2}{p^2}\right)  + \frac{Q^2 \left(7 t^4-6 t^2+1\right)
t^2}{64 p^2 \nu^4}\right\} \nonumber
\end{eqnarray*}
$p = z^2 + Q/2\nu t$. As before, we perform the calculations separately for the
contribution from each
power of $\nu$ using the formulas \Ref{AbelPlana} given above:
\begin{eqnarray*}
[\nu^0] &:& -\frac{2\cos\pi s}{\pi} (\mu R)^{2s} \sum_{l=1}\nu^{2-2s}
\int_0^\infty dz z^{2-2s}\frac{1}{p}\adb\\
&=&-\frac{2\cos\pi s}{\pi} (\mu R)^{2s} \int_0^\infty dz
z^{-2s}\sum_{l=1}\frac{\nu^{3-2s}}{\nu+a}\adb\\
&=&\left( -\frac{Q^3}{40\pi} - \frac{11Q}{48\pi}\right)\left[
\frac{1}{s} - 2 \ln \frac{\Omega}{2\mu} \right]  -
\frac{Q^3}{100\pi } + \mathcal{J}_1(Q) + \mathcal{J}_2(Q) +
\widetilde{\mathcal{J}}_1(Q),\\{}
[\nu^{-1}] &:& -\frac{2\cos\pi s}{\pi} (\mu R)^{2s} \sum_{l=1}\nu^{2-2s}
\int_0^\infty dz z^{2-2s}\frac{Qt}{4\nu p}\adb\\
&=&-Q\frac{\cos\pi s}{2\pi} (\mu R)^{2s} \int_0^\infty dz
z^{-2s} t \sum_{l=1}\frac{\nu^{2-2s}}{\nu+a}\adb\\
&= &\left(\frac{Q^3}{48\pi} + \frac{11Q}{96\pi}\right)\left[
\frac{1}{s} - 2\ln \frac{\Omega}{2\mu} \right]  +
\frac{Q^3}{72\pi }  - \frac{1}{2} \mathcal{J}_1(Q) +
\mathcal{J}_3(Q) + \widetilde{\mathcal{J}}_2(Q),\\{}
[\nu^{-3}] &:& -\frac{2\cos\pi s}{\pi} (\mu R)^{2s} \sum_{l=1}\nu^{2-2s}
\int_0^\infty dz z^{2-2s}\frac{Qt}{32\nu^3}\left[
\frac{t^2(35t^4-18t^2+1)}{p} +
\frac{2(7t^4-6t^2+1)}{p^2}\right]\adb\\
&=&-Q\frac{\cos\pi s}{16\pi} (\mu R)^{2s} \int_0^\infty dz z^{2-2s} t
\left[ \frac{t^2(35t^4-18t^2+1)}{z^2}\sum_{l=1}
\frac{\nu^{-2s}}{\nu+a} + \frac{2(7t^4-6t^2+1)}{z^4}
\sum_{l=1}\frac{\nu^{1-2s}}{(\nu+a)^2}\right]\adb\\
&=& \frac{23Q}{96\pi} \left[\frac{1}{s} -
\ln\frac{\Omega}{2\mu}\right] - \frac{547Q}{720\pi } + \mathcal{J}_4(Q) +
\mathcal{J}_5(Q)+ \widetilde{\mathcal{J}}_3(Q),\\{}
[\nu^{-4}] &:& -\frac{2\cos\pi s}{\pi} (\mu R)^{2s} \sum_{l=1}\nu^{2-2s}
\int_0^\infty dz z^{2-2s}\frac{Q^2t^2(7t^4-6t^2+1)}{64\nu^4
p^2}\adb\\
&=&-\frac{Q^2}{32\pi} \int_0^\infty dz z^{-2} t^2 (7t^4-6t^2+1)
\sum_{l=1} \frac{1}{(\nu+a)^2}\adb\\
&=& -\frac{11 Q}{240\pi} + \mathcal{J}_6(Q) +
\widetilde{\mathcal{J}}_4(Q),
\end{eqnarray*}
where  the following integrals were introduced,
\begin{eqnarray*}
\mathcal{J}_1(Q) &=& -\frac{Q^3}{16\pi} \int_0^\infty \frac{z
dz}{1+e^{\pi Q z/2}} \left[\frac{\textrm{arctanh}
\sqrt{1-z^2}}{\sqrt{1-z^2}} + \ln \frac{z}{2}\right],\adb\\
\mathcal{J}_2(Q) &=& \frac{2Q}{\pi}\int_0^\infty \frac{y^3
dy}{1+e^{2\pi y}} \int_0^\infty \frac{x^2 tdx}{x^4y^2 + Q^2 /4t^2}
\frac{1}{1+ x t},\adb\\
\mathcal{J}_3(Q) &=& \frac{Q}{\pi}\int_0^\infty \frac{y^3
dy}{1+e^{2\pi y}} \int_0^\infty \frac{x^3 t^2 dx}{x^4y^2 + Q^2/4t^2}
\frac{1}{1+x t},\adb\\
\mathcal{J}_4(Q) &=& \frac{Q}{8\pi} \int_0^\infty dx t^3 x^4
(35t^4-18t^2+1)\int_0^\infty \frac{1}{y^2x^4 + Q^2/4t^2}\frac{ydy}{1+e^{2\pi y}}
,\adb\\
\mathcal{J}_5(Q) &=& \frac{Q}{4\pi} \int_0^\infty dx t x^2
(7t^4-6t^2+1)\int_0^\infty \frac{x^4 y^2 - Q^2/4t^2 }{(y^2x^4 +
Q^2/4t^2)^2}\frac{y dy}{1+e^{2\pi y}},\adb\\
\mathcal{J}_6(Q) &=& \frac{Q}{16} \int_0^\infty dx t^3 x^4
(7t^4-6t^2+1)\int_0^\infty \frac{dy}{y^2x^4 +
Q^2/4t^2}\frac{y^2}{\cosh^2 \pi y},\adb\\
\widetilde{\mathcal{J}}_1(Q) &=& -\frac{Q}{2\pi} \int_0^\infty
\frac{x t dx}{(x^2 + Q/t)(x+Q)}\frac{1}{1 + xt},\adb\\
\widetilde{\mathcal{J}}_2(Q) &=& -\frac{Q^2}{4\pi} \int_0^\infty
\frac{dx}{x^2 + Q /t} + \frac{Q}{4\pi} \ln 2Q,\adb\\
\widetilde{\mathcal{J}}_3(Q) &=& \frac{Q}{8\pi} \int_0^\infty dx x^2
\left[\frac{t^3(35t^4 - 18t^2 +1)}{x^2 +  Q/t} + \frac{2t(7t^4 -
6t^2 +1)}{(x^2 + Q/t)^2}\right],\adb\\
\widetilde{\mathcal{J}}_4(Q) &=& \frac{Q^2}{8\pi} \int_0^\infty
\frac{dx t^2 x^2 (7t^4-6t^2+1)}{(x^2+Q/t)^2}.
\end{eqnarray*}
We note the following expressions which are necessary to consider the ideal conductor limit,
\begin{eqnarray*}
\lim_{Q\to \infty} \widetilde {\mathcal{J}}_2(Q) &=& -\frac{1}{8},\adb\\
\lim_{Q\to \infty}\widetilde{\mathcal{J}}_3(Q) &=& \frac{7}{256},\adb\\
\lim_{Q\to \infty}\widetilde{\mathcal{J}}_4(Q) &=& \frac{3}{256},
\end{eqnarray*}
All other integrals vanish in this limit.


\begin{thebibliography}{10}

\bibitem{abra70b}
M.~Abramowitz and I.A. Stegun.
\newblock {\em Handbook of Mathematical Functions}.
\newblock Dover, New York, 1970.

\bibitem{BI}
G.~Barton.
\newblock {Perturbative Casimir energies of dispersive spheres, cubes and
  cylinders}.
\newblock {\em J. Phys. A: Math. Gen.}, 34(19):4083--4114, 2001.

\bibitem{BIII}
G.~Barton.
\newblock Casimir energies of spherical plasma shells.
\newblock {\em J. Phys. A: Math. Gen.}, 37(3):1011--1049, 2004.

\bibitem{Blau:1988kv}
Steve Blau, Matt Visser, and Andreas Wipf.
\newblock {Zeta Functions and the Casimir Energy}.
\newblock {\em Nucl. Phys.}, B310:163, 1988.

\bibitem{BORDAG2006F}
M.~Bordag.
\newblock {Generalized Lifshitz formula for a cylindrical plasma sheet in front
  of a plane beyond proximity force approximation}.
\newblock {\em Phys.Rev.D}, 75:065003, 2007.

\bibitem{BORDAG2007B}
M.~Bordag.
\newblock On the interaction of a charge with a thin plasma sheet.
\newblock {\em Phys.Rev.D}, 76:065011, 2007.

\bibitem{Bordag:1999vs}
M.~Bordag, K.~Kirsten, and D.~Vassilevich.
\newblock {On the ground state energy for a penetrable sphere and for a
  dielectric ball}.
\newblock {\em Phys. Rev.}, D59:085011, 1999.

\bibitem{Bordag:2001qi}
M.~Bordag, U.~Mohideen, and V.~M. Mostepanenko.
\newblock {New developments in the Casimir effect}.
\newblock {\em Phys. Rep.}, 353:1--205, 2001.

\bibitem{Bordag:2005qv}
M.~Bordag, I.~G. Pirozhenko, and V.~V. Nesterenko.
\newblock Spectral analysis of a flat plasma sheet model.
\newblock {\em J. Phys.}, A38:11027, 2005.

\bibitem{Bordag:2004rx}
M.~Bordag and D.~V. Vassilevich.
\newblock Nonsmooth backgrounds in quantum field theory.
\newblock {\em Phys. Rev.}, D70:045003, 2004.

\bibitem{BOYER1968}
T.~H. Boyer.
\newblock {Quantum Electromagnetic Zero-Point Energy of a Conducting Spherical
  Shell and Casimir Model for a Charged Particle}.
\newblock {\em Phys.~Rev.}, 174:1764, 1968.

\bibitem{Graham:2003ib}
N.~Graham et~al.
\newblock {The Dirichlet Casimir problem}.
\newblock {\em Nucl. Phys.}, B677:379--404, 2004.

\bibitem{KN}
Mikhail~I. Katsnelson.
\newblock Graphene: carbon in two dimensions.
\newblock 2006.
\newblock arXiv:cond-mat/0612534.

\bibitem{Maksimenko2002}
S.A. Maksimenko and G.Ya. Slepyan.
\newblock {Electrodynamics of Carbon Nanotubes}.
\newblock {\em J. Comm. Tech. Electron.}, 47:235--252, 2002.

\bibitem{Milton:2004ya}
Kimball~A. Milton.
\newblock {The Casimir effect: Recent controversies and progress}.
\newblock {\em J. Phys.}, A37:R209, 2004.

\bibitem{Scandurra:1998xa}
Marco Scandurra.
\newblock The ground state energy of a massive scalar field in the background
  of a semi-transparent spherical shell.
\newblock {\em J. Phys. A: Math. Gen.}, 32:5679--5691, 1999.

\bibitem{Vassilevich:2003xt}
D.~V. Vassilevich.
\newblock {Heat kernel expansion: User's manual}.
\newblock {\em Phys. Rept.}, 388:279--360, 2003.

\end{thebibliography}
\end{document}